\documentclass[showpacs,prl,floatfix,twocolumn,amsmath,a4]{revtex4-1}
\usepackage{graphicx}

\begin{document}

\title{Magnetism and Charge ordering in TMTTF$_2$-PF$_6$ organic crystals}
  
\author{Gianluca Giovannetti$^{1}$, Sanjeev Kumar$^{2}$, Jean-Paul Pouget$^{3}$, Massimo Capone$^{1,4}$}

\affiliation{$^1$Democritos National Simulation Center, Consiglio Nazionale delle Ricerche, Istituto Officina dei Materiali (IOM) and Scuola Internazionale Superiore di Studi Avanzati (SISSA), Via Bonomea 265, 34136 Trieste, Italy}
\affiliation{$^2$Indian Institute of Science Education and Research Mohali, MGSIPAP Complex, Sector 26, Chandigarh 160 019, India}
\affiliation{$^3$Laboratoire de Physique des Solides, Universit\'e Paris Sud, CNRS, UMR 8502, 91405
Orsay, France}
\affiliation{$^4$Physics Department, University ``Sapienza", Piazzale A. Moro 2, 00185 Rome, Italy}
\date{\today}

\begin{abstract}
Using a combination of Density Functional Theory, mean-field analysis and exact diagonalization calculations
we reveal the emergence of a dimerized charge ordered state in TMTTF$_2$-PF$_6$ organic crystal. 
The interplay between charge and spin order leads to a rich phase diagram. Coexistence of charge ordering with
a structural dimerization results in a ferroelectric phase, which has been observed experimentally. The tendency to the dimerization is magnetically driven revealing TMTTF$_2$-PF$_6$ as a multiferroic material. 
\end{abstract}


\maketitle 

In recent years, organic molecular crystals have emerged as a fascinating class of materials with immense potential for applications. Structures in these materials are held together by van der Waals
and hydrogen bonding as opposed to much stronger covalent and ionic bonding in conventional crystals. As a consequence, these systems present complex phases and phenomena which still require a microscopic understanding, ranging from anomalous positive and negative thermal expansion\cite{Das} to high-temperature ferroelectricity with large polarization\cite{Horiuchi} and even superconductivity\cite{Mitsuhashi}. Moreover, their light, flexible and nontoxic character makes them ideal candidates for future applications.

The TMTTF$_2$-X series of charge transfer salts shows a very rich phase diagram when either hydrostatic or chemical pressure is applied \cite{Brazovskii,Jerome} with electronic phases ranging from spin/charge density waves to a spin-Peierls state. 
The case of TMTTF$_2$-PF$_6$ is peculiar. At ambient pressure it is a Mott insulator that develops a charge 
ordered (CO) state with a 4k$_F$ charge localization and the CO transition coincides with the onset
of ferroelectric order\cite{Monceau,Chow}. Moreover by lowering the temperature this state can be turned  into an antiferromagnetic (AFM) state or a Spin-Peierls phase.

In this paper we use a variety of theoretical and computational methods -- namely, density functional theory (DFT), 
mean-field (MF) and exact diagonalization (ED) -- to describe the low-temperature phase diagram and the emergence of a ferroelectric state in TMTTF$_2$-PF$_6$. We show that the interplay between charge-ordering, spin-ordering and structural dimerization leads to a rich phase diagram including a dimerized charge-ordered phase which breaks inversion symmetry and consequently is ferroelectric. 

We begin by presenting the results of DFT calculations, which are carried out 
using the generalized gradient approximation to the exchange-correlation according to Perdew-Becke-Ernzerhof (PBE) \cite{PBE}. 
The electronic structure of TMTTF$_2$-PF$_6$ is computed using Quantum Espresso package\cite{QE} and a tight-binding representation of the bandstructure is built using Wannier90 \cite{W90} to compute the maximally localized Wannier  orbitals.  The calculations for the low temperature centrosymmetric structure of TMTTF$_2$-PF$_6$ \cite{Granier} provide a simple picture: the HOMOs (highest occupied molecular orbitals) of the two TMTTF molecules hybridize, giving rise to bonding and antibonding bands around the Fermi level, with charge transfer to the PF$_6$ anions. 
The ratio 2:1 between cations (TMTTF molecules) and anions (PF$_6$ group) in the unit cell leads to a commensurate band filling of 3/4 of the TMTTF band. The system is metallic within PBE with a bandwidth W of $\sim$ 1.0 eV and it shows a quasi one-dimensional character along the stacking chain direction $a$ as reflected in the band structure.  The inability of PBE to reproduce the experimental insulating behavior is a very well known shortcoming of this approach when strong local and short-ranged correlations are relevant. 

\begin{figure}
\includegraphics[width=1.0\columnwidth,angle=0]{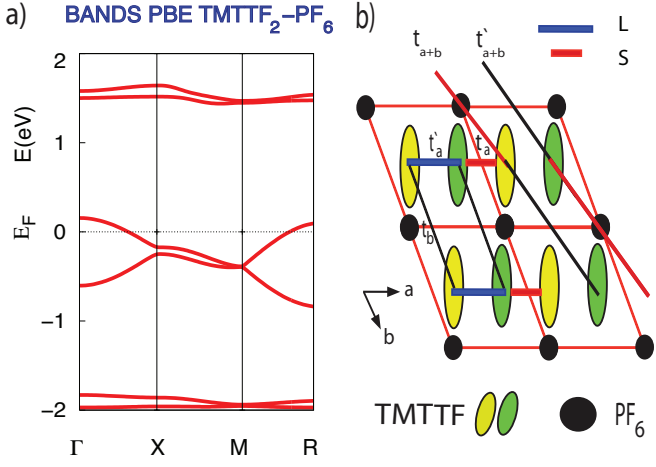}
\caption{(Color online) a) PBE band structure of TMTTF$_2$-PF$_6$ in the low temperature centrosymmetric phase. 
The high symmetry directions are labeled as: $\Gamma=(0,0,0)$, $X=(\frac{1}{2},0,0)$, $M=(\frac{1}{2},\frac{1}{2},0)$, 
$R=(0,\frac{1}{2},0)$. b) Schematic structure of various hopping parameters in TMTTF$_2$-PF$_6$ molecular crystals
in the $a$,$b$ plane along $a$ ($t_a$,${t^{'}}_a$) and $b$ ($t_b$), and the diagonal directions ($t_{a+b}$,${t^{'}}_{a+b}$).}
\label{fig1}
\end{figure}

In order to include such correlation effects we model TMTTF$_2$-PF$_6$ by a two-dimensional extended Hubbard model with the Hamiltonian:
\begin{eqnarray}
&& H = \sum_{i,j, \sigma} t_{i,j} [1 + \alpha (u_i - u_{i+1})] \left ( c^{\dagger}_{i \sigma} c^{~}_{j \sigma} + H.c. \right ) \nonumber \\
&& 
+ U \sum_{i} n_{i \uparrow} n_{i \downarrow} 
+ V \sum_{i,j} n_{i} n_{j} + V' \sum_{(i,j)} n_{i} n_{j} \nonumber \\ 
&& + \frac{1}{2} \sum_i u_i^2
\label{Ham}
\end{eqnarray}

Here, $c^{}_{i \sigma}$ and $c^{\dagger}_{i \sigma}$ are annihilation and creation operators for electrons at site $i$ with spin $\sigma$, $t_{ij}$ denote the bare hopping strengths and $u_i$ are the distortions along the stacking chain direction $a$ of the TMTTF ions from their equilibrium position.  The electron-lattice coupling parameter is $\alpha$, which favors dimerization. $U$ is the strength of the on-site Hubbard interaction, $V$ is the inter-site 
repulsion strength between TMTTF sites along $a$, and $V'$ is the inter-site 
repulsion strengths along $b$ and along the diagonal $a+b$ (see Fig. \ref{fig1}b)). 

The hopping parameters are extracted mapping the corresponding tight-binding model on the basis of maximally localized Wannier orbitals from PBE band structure. The only non-negligible hoppings are shown in Fig. \ref{fig1}b) and their values are $t_a$=0.22 eV, ${t^{'}}_a$=0.20 eV, $t_b$=0.04 eV, $t_{a+b}$=-0.01 eV, ${t^{'}}_{a+b}$=-0.03 eV, confirming the high anisotropy of the electronic properties.
These values agree well with experimental estimates of Ref. \onlinecite{Jacobsen} and are rather larger than those estimated on the basis of H\"uckel calculations \cite{Granier} in which the effect of the anions and of the screening of the crystal are not taken into account. The slight difference between the two hopping parameters along the $a$ axis $t_a$ and ${t^{'}}_a$ is due to 
the crystal arrangement of PF$_6$ anions which determines a dimerization of the TMTTF molecules (the two inequivalent distances S and L in Fig. \ref{fig1} have difference of $\sim$ 0.04 \AA \cite{Granier}). In terms of our model (\ref{Ham}), this corresponds to ${\alpha}\sim$1.2 assuming $t_a$=${t^{'}}_a$=0.21 eV.

The Hubbard $U$ is the screened energy cost required to doubly occupy a molecular site with holes. This energy cost can be evaluated explicitly from the total energy of an ion with different number of electrons and the polarization energy \cite{Ueff}. We calculate the variation 
of the eigenvalues and their polarization energy of the (hole) charged molecules placed inside a cavity of an 
homogeneous dielectric medium with dielectric constant of 3 using the SS(V)PE model \cite{Chipman00}. 
We estimate $U$=2.2 eV using the GAMESS package \cite{GAMESS} using TZV basis set and PBE functional.  

Since little is known theoretically about the inter-site interaction parameters $V$,$V'$ we adopt a phenomenological approach  studying the phase diagram letting these parameters vary in the typical range of other molecular crystals \cite{Cortes}. An experimental estimate $V \simeq 2t$ \cite{Mila} would imply $V \simeq$ 0.4 eV, which lies in the middle of the range we consider.

Employing a mean-field Hartree-Fock (MF) decomposition for the Hubbard U and $V$,$V'$ terms, we solve the model in Eq. \ref{Ham} with the hopping parameters listed above self-consistently for 
different initial conditions and retain the lowest-energy solutions using a unit cell of 4$\times$4 sites. 
\begin{figure}
\includegraphics[width=1.0\columnwidth,angle=0]{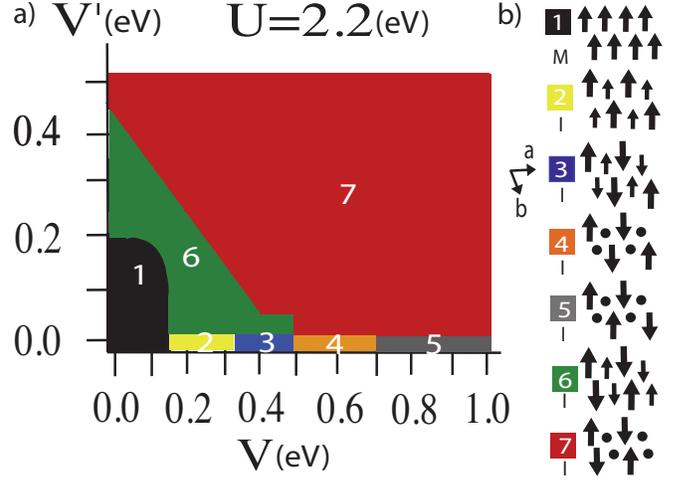}
\caption{(Color online) (a) Phase diagram showing the magnetic and charge ordered phases in the parameter space of  $V$ and $V'$ (eV). M and I label Metallic and Insulating ground states respectively. A schematic view of charge and  magnetic orders appearing in the phase diagram is shown in (b) for a unit cell 4x2.}
\label{fig3}
\end{figure}
In Fig. \ref{fig3}, we show the groundstate phase diagram in the parameter space of $V$ and $V'$. 
We obtain a variety of phases that emerge from the interplay between the on-site and inter-site repulsions. The main tendency at MF level is the charge ordering which indeed occurs at any finite values of $V$ and $V'$. We find essentially two structures: (i) the CO pattern with wave vector q$=(\pi,0)$ (phases 6, 7 in Fig. \ref{fig3}), stabilized  with increasing $V$ at finite $V'$, which can be directly connected with the ferroelectric state for X=PF$_6$, and (ii) the CO pattern with wave vector q$=(\pi,\pi)$ (phases 2, 3, 4, 5 in Fig. \ref{fig3}), stabilized with increasing $V'$ at $V$=0, which is similar to the antiferroelectric state observed with X=SCN \cite{Coulon}.

To a large extent, the spin-ordering follows the CO pattern and combines with it leading to the rich phase diagram we have drawn. For small values of V the ferromagnetic (FM) order $\Uparrow$-$\Uparrow$-$\Uparrow$-$\Uparrow$ along the stacking chain direction $a$ is the groundstate (phase 1 in Fig. \ref{fig3}) 
with a lower energy with respect  to the AFM state $\Uparrow$-$\Uparrow$-$\Downarrow$-$\Downarrow$ which is realized without CO at higher energy. 
The FM solution can be tuned into a CO checkerboard state at finite $V$ and $V'$=0 (phase 2 in Fig. \ref{fig3}).
Increasing $V$ at finite $V'$ an AFM state with charge and spin modulation along direction $a$ establishes. In such state the spin pattern is 
$\Uparrow$-$\uparrow$-$\Downarrow$-$\downarrow$ (where the single arrows refer to a small magnetic moment and the double arrows to a large one) and staggered charge pattern ${\delta}^+$-${\delta}^-$-${\delta}^+$-${\delta}^-$ (${\delta}^{\pm}$ indicating a positive or negative charge imbalance with respect to the average) leading to phases 3 and 6. 
At larger $V$ the AFM state turns into the extreme picture $\Uparrow$-$O$-$\Downarrow$-$O$ with alternating singly and doubly occupied TMTTF sites, the latter having zero magnetic 
momentum (phases 4, 5 and 7). It is important to realize that in our two-dimensional model the behavior along the $a$ axis combines with  the CO pattern along the b direction that we discussed above leading to five different phases (numbered from 3 to 7). All these phases are characterized by 2k$_F$ spin ordering and 4k$_F$ CO along the $a$-direction. They are consistent with the experimental evidence of spin \cite{Nakamura} 
and charge \cite{Monceau} ordering in TMTTF salts and well agree also with calculations in the one-dimensional extended Hubbard model \cite{Seo}. Our calculations confirm these 
results in a more realistic two-dimensional model, and show that it is of great importance to take into account the complicated details of the interchain couplings to study the competition of the different phases.
The degree of anisotropy of the intersite interactions in TMTTF2-PF$_6$ crystals is more complicated than the one used in our model however our approach captures both CO patterns revealed from experiments in this class of materials.

Lattice effects can accompany the CO phase \cite{Souza} and indeed a 4k$_F$ bond dimerization has been reported experimentally for the CO phase of TMTTF$_2$-PF$_6$ \cite{Granier}.
Atomic displacements, which break the inversion symmetry, can give rise to ferroelectricity once they are coupled with a coherent charge order as in the case of TMTTF$_2$-PF$_6$.
Here in the following we discuss, as a natural generalization of the dimerized hopping structure due to the inequivalent distribution of PF$_6$ anions assumed so far, the possibility of  formation of dimers between TMTTF sites along the stacking chain direction $a$ as consequence of minimization of the exchange energy: the onset of AFM ordering pins the dimer formation tendency.
For the sake of definiteness, we assume in the following $V'=\frac{1}{2}V$ (obviously $V'$ is expected to be smaller than $V$ because of the larger distance involved).

\begin{figure}
\includegraphics[width=1.0\columnwidth,angle=0]{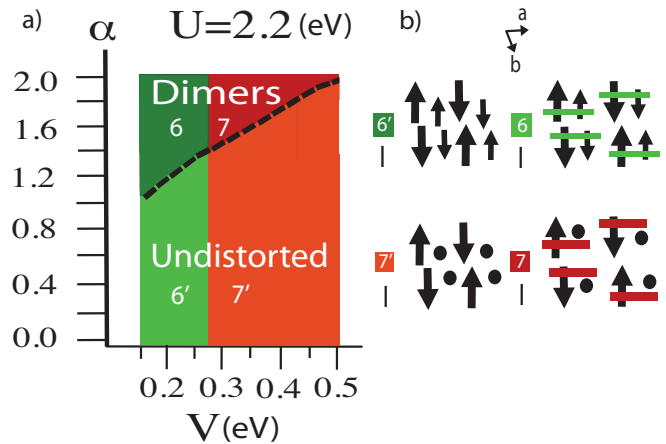}
\caption{(Color online) Phase diagram showing the dimerization of magnetic and charge order phases in the parameter space 
of $\alpha$ and V. Here we have used $V' = V/2$. }
\label{fig4}
\end{figure}

We now turn to the model of Eq. \ref{Ham} considering a symmetrized hopping ($t_a = {t^{'}}_a$=0.21 eV) and we vary $\alpha$ from 0 to 2 in order to explore the interplay between CO and dimerization. The energy is minimized as a function of the classical distortions $u_i$.
In Fig. \ref{fig4} the phase diagram shows a transition from an undistorted state (phases $6'$ and $7'$) to a dimerized state (phases 6 and 7) with increasing $\alpha$.  The key point is that a larger $\alpha$ is needed to stabilize dimerization when CO is stronger.
The value of $\alpha$ required to favor the transition to a dimerized state is in good agreement with our DFT estimate in the whole range of $V$'s. Our results complement previous similar reports \cite{YOSHIOKA,Riera,Mazumdar} within a realistic framework in which the bandstructure is derived from ab-initio calculations.

The 4k$_F$-CDW of a Wigner type with dimerization of TMTTF lacks inversion symmetry and is therefore ferroelectric as experimentally reported \cite{Monceau}. 
Equivalent displacements (${\Delta}s$) of molecular sites in the formation of dimers along the stacking direction with disproportionation of charge (${\delta}^+$-${\delta}^-$) create local dipoles of value 2${\delta}$${\Delta}s$ at each dimer, which sum up along the lattice in the $a$ direction resulting in a finite polarization in the ferroelectric 
CO at ($\pi$,0). In this scenario the anion ${{PF}_6}^-$ is expected to move closer to the molecular site with less electrons (${\delta}^+$) making the CO stronger \cite{Foury}.
The CO states (phases 6, 7 in Fig. \ref{fig3}) are characterized by a unit cell containing an essentially singly occupied TMTTF molecule (A) and an essentially doubly occupied one (B). In these states the spin-minority states at site A are empty and can be occupied by minority electrons jumping from site B creating exchange striction of the same kind found in novel organic multiferroic materials \cite{Mostovoy}. The stronger the magnetic coupling, the easier it is for the system to dimerize. Therefore the magnetism is controlling the ferroelectricity. 
The symmetry of the insulating state is such that magnetic ordering pushes a charge-ordering pattern from site-centered towards bond-centered \cite{Efrimov}.  
In this intuitive picture of the origin of the ferroic state we treat electrons as localized however the covalency 
of the electronic state pumps the charge through the crystal by quantum effects \cite{Nagaosa}.  
It is interesting to observe that moderate values of $V \sim$ 0.5eV are sufficient to lead to this extreme charge disproportion despite the larger value of $U$.
 
Here we remark the importance of using a full two-dimensional modeling to properly account for the magnetic and charge properties. A one-dimensional description would indeed miss the competition of anti-ferroelectric phases, whose occurrence depends on complicated details of the interchain coupling.
The present MF scheme however does not take into account quantum fluctuations, and hence overestimates the stability of  magnetic and charge orderings.
\begin{figure}
\includegraphics[width=1.0\columnwidth,angle=0]{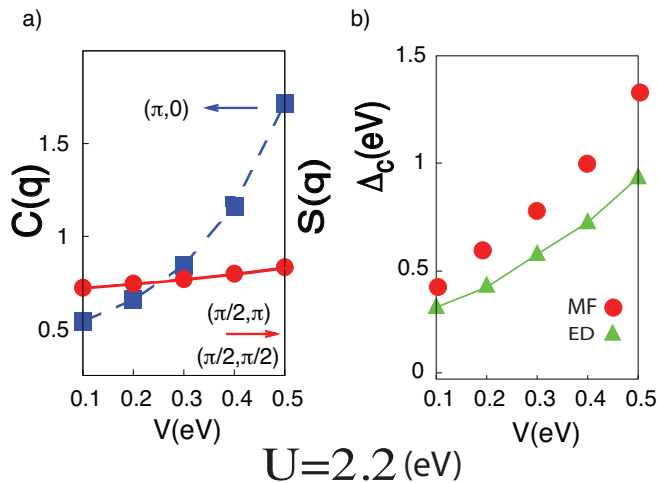}
\caption{(Color online) (a) Spin and charge structure factors $S(q)$ and $C(q)$ at some relevant wave vectors (The two $S(q)$ for different momenta lie on top of each other). (b) Charge gap  ${\Delta}_c$ at different values of V and $V' = V/2$ in MF and ED schemes.}
\label{fig5}
\end{figure}
In order to take into account such fluctuations, we solve (\ref{Ham}) by Lanczos ED on a 3/4-filled finite square cluster of 4$\times$4 sites (N) with the same parameters of our MF analysis. Within this approach we calculate the charge gap (${\Delta}_c$)  and the structure factors $\frac{1}{N} \sum_i \exp(iqr_i) (n_{i\uparrow} \pm n_{i\downarrow})$, where the plus and minus sign correspond to the charge ($C(q)$) and spin ($S(q)$) structure factor respectively\cite{Dagottoreview}. 

In Fig. \ref{fig5} we show $C(q)$ and $S(q)$ for the momenta that maximize the response functions as a function of $V$. The ED $C(q)$ confirms indeed that $V$ favors CO with wave vector q=(${\pi}$,0) (4k$_F$) for V ${\geq}$ 0.2 eV (see Fig. \ref{fig5}a)), while for small $V$ it is competitive with the q=(${\pi}$,${\pi}$) CO (not shown). 
$S(q)$ displays a smaller enhancement as a function of $V$ and two magnetic orderings are emphasized: (i) an AFM state with wave vector q=($\frac{\pi}{2}$,$\pi$) ($\Uparrow$-$\Uparrow$-$\Downarrow$-$\Downarrow$) \cite{notaSq} and 
(ii) another AFM state with wave vector q=($\frac{\pi}{2}$,$\frac{\pi}{2}$). 
The two magnetic orderings are consistent with the one we find in MF for X=PF$_6$ and observed experimentally for X=SbF$_6$ and X=SCN \cite{Nakamura} respectively.
Coupling charge and magnetic states we can recover the same phases 6 and 7 obtained in MF.
The charge gap ${\Delta}_c$ is calculated averaging over twisted boundary conditions \cite{Koretsune} in order to reduce finite-size effects and the MF values are shown in Fig. \ref{fig5}b) for selected values of $V$. The  charge gap ${\Delta}_c$ is finite over the entire range of $V$. Increasing the lattice size is expected to reduce the value of the charge gap moving toward the experimental value \cite{Laversanne} and suggesting that a small value of V may be relevant for such organic structures. Moreover the temperature effects not taken into account in the present study soften both spin and charge orders.  

The tendency toward formation of dimers along the stacking direction $a$ is qualitatively maintained within ED calculations.
The competition between two insulating phases may lead to a metallic state in the $V$-$V^{'}$ phase diagram which results from the geometrical frustration of the long-range Coulomb interactions \cite{Merino} but due to the anisotropy and large on-site Coulomb repulsion the TMTTF$_2$-PF$_6$ salt is in a dimer Mott insulating state. 
In conclusions with a combination of different techniques we have highlighted a peculiar novel interplay between charge, spin and lattice degrees of freedom in the low temperature phase diagram of TMTTF$_2$-PF$_6$ crystals.
The magnetic and charge orderings coming out from MF and ED analysis are consistent and agree with experiments suggesting that the competition between local and non-local Coulomb repulsion is the key to understand the properties of this compound. The charge ordered state in conjunction with the spin ordering makes the system a natural candidate for being a new multiferroic material.

We acknowledge useful discussions with D. Khomskii and N. Nagaosa. GG and MC acknowledge financial support by European Research Council under FP7/ERC Starting Independent Research Grant ``SUPERBAD" (Grant Agreement n. 240524). 
Computational support from CINECA is gratefully acknowledged.


\begin{thebibliography}{100}
\bibitem{notaSq} The magnetic orders with propagation vectors q=($\frac{\pi}{2}$,$\pi$) and =($\frac{\pi}{2}$,0) are related by the symmetry of the lattice so that a change in the setting of the lattice vectors result in the same spin arrangement.
\end{thebibliography}
\end{document}